\documentclass[12pt]{article}

\usepackage{sbc-template}
\usepackage{graphicx,url}
\usepackage[latin1]{inputenc}
\usepackage{amsmath}
\usepackage{amsfonts}
\usepackage{amssymb}
\usepackage[usenames]{color}

\newcommand{\bra}[1]{\langle{#1}\vert}
\newcommand{\ket}[1]{\vert{#1}\rangle}
\newcommand{\braket}[2]{\langle{#1}\vert{#2}\rangle}

\def\C{\mathbb{C}}

\def\H{\mathcal{H}}

\title{Decoherence in Search Algorithms}
\author{
        G.~Abal\inst{1},
        R.~Donangelo\inst{1},
        F.L.~Marquezino\inst{2},
        A.C.~Oliveira\inst{3},
        R.~Portugal\inst{2}}
\address{
Instituto de F\'\i sica, Universidad de la Rep\'ublica, \\ %
Casilla de Correo 30, C\'odigo Postal 11300, Montevideo, Uruguay %
\nextinstitute %
Laborat\'orio Nacional de Computa\c c\~ao Cient\'{\i}fica,\\ %
Avenida Get\'ulio Vargas, 333, Petr\'opolis, RJ, 25651-075, Brazil %
\nextinstitute %
Instituto de Ci\^encias Exatas, Universidade Federal de Lavras, \\ %
Campus Universit\'{a}rio, Lavras, MG, 37200-000, Brazil %
\email{abal@fing.edu.uy,donangel@if.ufrj.br,amanda2711@gmail.com}\email{\{franklin,portugal\}@lncc.br}
 }

\begin{document}

\maketitle

\begin{abstract}
Recently several quantum search algorithms based on quantum walks were proposed. Those algorithms differ from Grover's algorithm in many aspects. The goal is to find a marked vertex in a graph faster than classical algorithms. Since the implementation of those new algorithms in quantum computers or in other quantum devices is error-prone, it is important to analyze their robustness under decoherence. In this work we analyze the impact of decoherence on quantum search algorithms implemented on two-dimensional grids and on hypercubes.
\end{abstract}

\begin{resumo}
Recentemente, v\'{a}rios algoritmos qu\^{a}nticos de busca baseados em
passeios aleat\'{o}rios qu\^{a}nticos foram apresentados. Estes algoritmos s\~{a}o diferentes
do algoritmo de Grover em v\'{a}rios aspectos. O objetivo \'{e} encontrar um
v\'{e}rtice marcado em um grafo mais r\'{a}pido do que algoritmos cl\'{a}ssicos. Uma vez
que a implementa\c{c}\~{a}o destes novos algoritmos em computadores qu\^{a}nticos ou
qualquer outro dispositivo qu\^{a}ntico est\'{a} sujeito a erros, \'{e} importante analisar a
robustez em rela\c{c}\~{a}o \`{a} descoer\^{e}ncia. Neste trabalho, analisamos a descoer\^{e}ncia
dos algoritmos qu\^{a}nticos de busca em malhas bi-dimensionais e hipercubos.
\end{resumo}

\section{Introduction}

After Grover's seminal paper \cite{Gro96a}, it is known that a
quantum computer can search an element in a non-structured
database quadratically faster compared to a classical computer.
Recently many other search algorithms were developed based on the
discrete-time quantum walk model \cite{Kempe03}. They differ from
Grover's algorithm in many aspects and may be better suited for
practical implementation \cite{Shenvi,AKR}.

Quantum walks can be seen as a quantum counterpart of
classical random walks \cite{Kempe03}. They have been successfully
applied in several quantum algorithms
\cite{Amb03,Szegedy04,mss07}. It is possible to describe an
\textit{abstract search algorithm}  \cite{AKR} in a generic regular graph by
using a discrete-time quantum walk with a modified coin operator
to search for a marked vertex. Grover's algorithm can
be seen as the simplest example of this procedure. The
Shenvi-Kempe-Whaley (SKW) algorithm can be seen as an implementation of the
abstract search algorithm to an $n-$dimensional
hypercube \cite{Shenvi} and the Ambainis-Kempe-\-Rivosh (AKR)
algorithm is an application to two-dimensional grids \cite{AKR}.

Decoherence and gate imperfections are unavoidable side-effects in
any implementation of a quantum computer. One approach to deal
with this problem, with considerable overhead in quantum
resources, consists in using redundant encoding and several layers
of error-correction codes. Another approach consists in designing
algorithms that are robust against certain types of errors which
may be dominant in a given implementation. This requires a
detailed knowledge of the effect that different kinds of noise
have on the performance of the algorithm. It seems likely that a
real quantum computer will take advantage of both approaches.
Decoherence in quantum walks has been considered in previous
works, see for instance \cite{KT02,deco,Alagic}.

In this work, we study the effect of three different noise models
on the AKR and SKW algorithms without quantum error correction. In two of them we assume phase errors affecting the modified coin operator. In the other case, we assume broken-link
imperfections affecting the shift operator. The influence of
errors in the AKR algorithm has not been analyzed before. This
work presents the first results in this direction. Li and
collaborators \cite{Li2006} have analyzed the effect on the SKW
algorithm of gate errors in the coin operator assuming a perfect
shift operator. In this work we complement those results by analyzing errors in the shift
operator. 

The structure of the paper is as follows. In Section~2, we review
Grover's algorithm and introduce the notation used in this work.
In Section~3 we review some basic concepts of search algorithms
based on discrete-time quantum walks. In Section~4, we describe
the noise models studied in this work and review the
broken-link type of noise \cite{Mixing}. In Section~5, we review
the SKW Algorithm. In Section~6 we review the AKR Algorithm. In
Section~7, we describe the methodology used to study each of
the noise models and give the results of the numerical
simulations we performed for the SKW algorithm. In Section~8, we
present the results of the numerical simulations for the AKR
algorithm. In Section~9, we present our conclusions.

\section{Grover's Algorithm}

In this section we briefly review Grover's algorithm \cite{Gro96a}
and the notation used in this paper. For further details see
\cite{NC00}. Consider a quantum computer of $n$ quantum bits
(qubits). Quantum Mechanics tells us that its state is described
by a unit vector in a vector space of dimension $N=2^n$. The
simplest orthonormal basis for this vector space is
$\{\ket{0},\ket{1},\cdots,\ket{N-1}\}$, where $\ket{i}$ is a
vector of $N$ entries all of them zero except the entry $i+1$
which is 1. This basis is called \textit{computational basis}.
Suppose that the state of the quantum computer in a given instant
of time is the vector $\ket{\psi}$, then
\begin{equation} \label{eq1}
\ket{\psi}=\sum_{i=0}^{N-1} \alpha_i \ket{i}
\end{equation}
where the coefficients $\alpha_i$ are complex numbers that must
obey the constraint
\begin{equation} \label{eq2}
 \sum_{i=0}^{N-1} |\alpha_i|^2 = 1.
\end{equation}
It is possible to prepare the quantum computer at the beginning of
an algorithm in any state $\ket{i}$. The algorithm must be a
sequence of applications (multiplications) of $N\times N$ unitary
operators (matrices) $U_1,\cdots,U_t$. So, at time $t$, the state
of the quantum computer is
\begin{equation} \label{eq3}
\ket{\psi}=U_t \cdots U_1 \ket{i}.
\end{equation}
A matrix $U$ is unitary if $U U^\dagger = I$, where $U^\dagger$ is
the transpose conjugate of $U$ and $I$ is the identity matrix.

Since the quantum computer is a physical system, one can perform
measurements to determine the state $\ket{\psi}$. Quantum
Mechanics tells us that the result of a complete measurement in
the computational basis will be $i$ with probability
$|\alpha_i|^2$. The measurement does not allow one to find the
coefficients $\alpha_i$, that are needed to describe the vector
$\ket{\psi}$. Instead, one gets a random number in the set
$\{0,\cdots,N-1\}$ with the corresponding probability distribution
$\{|\alpha_0|^2,\cdots,|\alpha_{N-1}|^2\}$. A complete measurement
means that all qubits are measured yielding either 0 or 1 each.
One gets a binary number that is converted to the decimal
notation. A partial measurement consists in measuring a fraction
of the qubits.

A search algorithm, such as Grover's algorithm, aims to determine
whether an element $i_0$ belongs to non-structured database or to
determine the position of an element in a non-sorted database. The
easiest way to pose this problem is in the following
form. Suppose that the domain of function $f$ is
$\{0,\cdots,N-1\}$ and the image is
\begin{equation} \label{f}
f(x)=
  \begin{cases}
    1 & \text{if }{x=i_0}, \\
    0 & \text{otherwise}.
  \end{cases}
\end{equation}
Suppose that we ask a friend to implement the function $f$ in a
classical computer and there is only one value of $x$ such that
$f(x)=1$. We can obtain the image of any input value by employing
$f$. What is the complexity of the best algorithm that finds the
value of $x$ such that $f(x)=1$? The complexity in this case is
measured by the number of times we employ the function $f$. If we do
not know any equation for $f$ nor any details of the
implementation of $f$, the only way to find out that $f(i_0)=1$ is
through an exhaustive search. The complexity of the best classical algorithm
is $O(N)$.

In a quantum computer, the implementation of function $f$ must be
performed through a unitary matrix, which we call $U_f$. The
definition of $U_f$ is
\begin{equation} \label{Uf}
U_f \ket{x}\ket{0}=
  \begin{cases}
    \ket{i_0}\ket{1} & \text{if }{x=i_0}, \\
    \ket{x}\ket{0} & \text{otherwise}.
  \end{cases}
\end{equation}
We suppose that the quantum computer has $n+1$ qubits. So the
output of the function $f$ is added to a second register of 1
qubit. We again ask our friend to implement $U_f$, this time in a
quantum computer. What is the complexity of the best quantum
algorithm that finds the value of $x=i_0$? Grover's algorithm
requires $O(\sqrt{N})$ applications of $U_f$ to determine the
value $i_0$ with a very small margin of error.

The idea behind the algorithm is to start the quantum computer in
a known state, then apply a sequence of unitary matrices that
results in a state that has a large overlap with state
$\ket{i_0}$. If the state of the quantum computer $\ket{\psi}$ has
a large overlap with $\ket{i_0}$, that is
$\braket{\psi}{i_0}\approx 1$, the result of the measurement will
be $i_0$ with high probability. The notation $\bra{\psi}$ means
the transpose conjugate of $\ket{\psi}$.

Grover's algorithm works as follows. Prepare a $(n+1)$-qubit
quantum computer in state $\ket{\psi_0}=\ket{0}\ket{1}$. Apply
$H^{\otimes (n+1)}$ to $\ket{\psi_0}$. $H$ is the Hadamard matrix
and $\otimes$ is the tensor product, see \cite{NC00}. The
result is $\ket{\psi_1}= \ket{s}\ket{-}$, where
$\ket{-}=(\ket{0}-\ket{1})/\sqrt{2}$ and
\begin{equation}
\ket{s}=\frac{1}{\sqrt{N}}\sum_{i=0}^{N-1}\ket{i}.
\end{equation} \label{U_Grover}
Now apply $U^{\lceil\frac{\pi}{4} \sqrt{N}\rceil}$ to
$\ket{\psi_1}$, where
$$U=((2\ket{s}\bra{s}-I_{N})\otimes I_{2})\ U_f.$$ %
In the last step, measure the first register to obtain $i_0$ with
probability $1-O(1/{N})$.

\section{Quantum Walk based Search Algorithms}

Quantum walks generalize the concept of classical random walk. A
classical walk is a prescription of how to move,
conditioned to the value of a random variable. If the walker
lives in a regular graph of degree $d$, the random variable must
have $d$ values, usually with the same probability $1/d$ (a balanced walk). The
edges of the graph incident to a vertex $v$ must have labels from
0 to $d-1$. If the walker is in vertex $v$ and the result of the
random variable is $j$, than the walker moves to the vertex
$v^\prime$ that is connected to $v$ by an edge of label $j$. This
procedure is repeated again and again. The result is a random walk
on the graph. In a one-dimensional lattice, one can toss a coin and
move to the right if the result is heads, or to the left if the
result is tails.

In a quantum setting, both the toss of a coin and the shift of the
walker must be performed by unitary operators. In a regular graph
of degree $d$, the vector space where the walk takes place is
$\mathcal{H}_C\otimes\mathcal{H}_V$, where $\mathcal{H}_C$ is the
Hilbert space spanned by $\{\ket{0},\cdots,\ket{d-1}\}$
representing the coin space and $\mathcal{H}_V$ is the Hilbert
space spanned by $\{\ket{0},\cdots,\ket{V-1}\}$ representing the
vertex space, where $V$ is the number of vertices. The usual form
of the evolution
operator is %
\begin{equation}U = S\ (C\otimes I).\end{equation}%
Here, $C$ is a $d\times d$ matrix that acts only on the coin space, $I$ is the
identity in the vertex space and
$S$ is the shift operator given by %
\begin{equation}S\ket{j}\ket{v}=\ket{j}\ket{v^\prime},   \end{equation}
where $v^\prime$ is the vertex that is connected to $v$ by edge
$j$. Note that the coin operator $C$ is the same for all vertices. The
walker starts at some initial configuration called $\ket{\psi_0}$
and at time $t$ its state is $U^t\ket{\psi_0}$.

Quantum walk search algorithms are based on a modification of some
standard quantum walk given by $U$. The analysis of the new walk
depends on what happens in the original non-modified walk. Suppose
that we would like to search for vertex $v_0$. That vertex must be
marked somehow. One marks this vertex using a modified coin
operator. The new coin operator must distinguish the marked vertex
from the rest
The new coin is defined by %
\begin{equation} \label{Cprime}
C^\prime=(-I)\otimes\ket{v_0}\bra{v_0}+C\otimes(I-\ket{v_0}\bra{v_0}).
\end{equation}%
This new coin operation $C^\prime$ defines a new evolution operator given by
$U^\prime=S\ C^\prime$.

The most used coin is called Grover's coin and it is the real
unitary operator farthest from the identity \cite{Moore}. It is
defined as
$C=2\ket{s}\bra{s}-I$, where $\ket{s}$ is the uniform superposition, %
\begin{equation} \label{eq_s}
\ket{s}=\frac{1}{\sqrt{d}}\sum_{i=0}^{d-1}\ket{i}.
\end{equation}
For this coin, all directions have the same weight. Using Grover's
coin in Eq.~(\ref{Cprime}), one obtains %
\begin{equation} \label{Uprime}
U^\prime=U\ (I-2\ket{s,v_0}\bra{s,v_0}).
\end{equation}
It is possible to perform a very general analysis of search
algorithms on graphs if one demands some properties from
$U$ \cite{AKR}. These properties are: (1) $U$ must be a real
unitary matrix, (2) $U$ must have only one eigenvector with
eigenvalue 1, and (3) this eigenvector must be the initial state
of the algorithm. This \textit{abstract search algorithm} works as
follows. Suppose that $U\ket{\phi_0}=\ket{\phi_0}$ and let $\exp(i
\alpha)$ be the eigenvalue of $U^\prime$ with the smallest angle
among all eigenvalues of $U^\prime$. If $U^\prime$ has eigenvalue 1, the initial
condition and the evolution of the walk must be in a space orthogonal to the eigenspace
associated with eigenvalue 1.
The algorithm consists in applying
$U^{\prime\ \lceil\frac{\pi}{2\alpha}\rceil}$ to $\ket{\phi_0}$
and measuring the vertex space.

Grover's algorithm is the simplest example of form (\ref{Uprime}).
Two new search algorithms in this framework have been analyzed in
detail. The first one is a search on hypercubes (SKW), which we
describe in Sec.~\ref{sec:SKW}. The second one is a search on two
dimensional grids (AKR), which we describe in Sec.~\ref{sec:AKR}.

\section{Decoherence models}

In an actual physical implementation, operators are error-prone.
It is important to determine the robustness of an
algorithm to errors in its implementation. In quantum search
algorithms, there are three key operators: the original coin
($C$), the coin used in the marked vertex ($-I$) and the shift
($S$). We analyze the impact that errors on each of these operators have on the
algorithm's performance.

Phase errors on the coin operator affecting the marked node can be
implemented by replacing $-I$ by
\begin{equation}\label{Cv0}
\widetilde{C}_{v_0}(\theta) = e^{i(\pi + \theta)} I
\end{equation}
with $\theta\in [-\pi,\pi]$. The perfect coin operator $-I$ on the
marked vertex is recovered for $\theta = 0$. We say that the error
is systematic when the phase error $\theta$ is constant in each
step (model~I), and that the error is random when the phase error
$\theta$ in each step is a Gaussian random variable with zero mean
and standard deviation $\sigma$ (model~II).

Phase errors on the coin operator for the unmarked nodes can be
implemented by rewriting $C$ as
\begin{equation}
\widetilde{C}(\theta) = I - (1-e^{i(\pi + \theta)}) \ket{s}\bra{s}
\end{equation}
with $\theta\in [-\pi,\pi]$. The Grover coin operator is recovered
for $\theta = 0$.

The effect of phase errors in the original Grover's algorithm was
analyzed by Long and collaborators and, later on by Shenvi and
collaborators \cite{Long00,Shenvi-noise}. The latter authors investigated
the importance of the scaling of phase errors with the size of
the database, $N$. In a recent work, Li and
collaborators \cite{Li2006} studied the effect of an imperfect $C$
on the SKW algorithm. The operators $-I$ (acting on the marked vertex) and
$S$ were supposed to have no errors.

Errors in the shift operator $S$ can be implemented by randomly
opening links between connected vertices with probability $p$ per
unit time (model~III). This broken-link noise model has been
previously considered for a quantum walk on a line and on
a plane \cite{deco,Amanda}, and on the hypercube  \cite{Mixing}. To
implement this kind of error we generalize the shift operator $S$
such that no probability flux is transferred across a broken link.
This modified shift operator is unitary for any number of broken
links in the lattice. At each time step, the topology of the graph
is defined, opening each link with probability $p$ and performing
the shift to the neighboring vertex only if the link is not
broken. The original $S$ operator is recovered for $p=0$.

\section{SKW Algorithm}\label{sec:SKW}

The quantum search on $n$-dimensional hypercubes has a Hilbert
space $\H_C \otimes \H_P$, where $H_C$ is a $n$-dimensional
Hilbert space associated with a ``quantum coin'' and $\H_P$ is a
$2^n$-dimensional Hilbert space associated with the vertices of
the hypercube. A basis for $\H_C$ is $\{\ket{d}\}$, for $0 \leq d
\leq n-1$ and a basis for $\H_P$ is $\{\ket{x_{n-1}x_{n-2}\cdots
x_0}\}$, for binary $x_d$. In a hypercube two vertices are
connected if, and only if, the corresponding binary strings differ
by one bit.

The generic state of the quantum walker is
\begin{equation}
\ket{\Psi(t)} = \sum_{d=0}^{n-1}\sum_{\vec{x}=0}^{2^n} \psi_{d;\vec{x}}(t) \ket{d,\vec{x}},
\end{equation}
where $\psi_{d;\vec{x}}(t) \in \C$ and
$\sum_{d=0}^{n-1}\sum_{\vec{x}} |\psi_{d;\vec{x}}(t)|^2 = 1$. The
evolution operator for one step of the walk is $U=S\ (C\otimes
I)$, where $S$ is the shift operator and $C$ the coin operator
acting in $\H_C$ and $I$ is the identity in $\H_P$. The shift
operator can be written as
\begin{equation}
S = \sum_{d=0}^{n-1}\sum_{\vec{x}}\ket{d,\vec{x}\oplus \vec{e}_d}\bra{d,\vec{x}},
\label{eq:S}
\end{equation}
where $\oplus$ is the bitwise binary sum between binary vectors,
and $\vec{e}_d$ is a null vector except for a single $1$ entry in
the $d$-th component. The coin operator is given by
\begin{equation}
C =  2\ket{s^C}\bra{s^C} - I, \label{eq:grover-coin}
\end{equation}
where $\protect{\ket{s^C} =
\frac{1}{\sqrt{n}}\sum_{d=0}^{n-1}\ket{d}}$ is the uniform
superposition over all directions.

In the SKW search algorithm, the initial state is the uniform
superposition of coin and spatial eigenstates, {\it i.e.},
$\ket{\psi_0} = \ket{s^C}\otimes \ket{s^P}$, where $\ket{s^P}$ is
the uniform  superposition over the $2^n$ nodes of the hypercube.
Note that a single searched state is contained in $\ket{s^P}$ with an amplitude $1/\sqrt{2^{n}}$.  The initial state
$\ket{\psi_0}$ is an eigenstate of $U$ with eigenvalue 1. The quantum search (SKW), as proposed
by \cite{Shenvi}, is based on a modified coin operator of the form
\begin{equation}
C^\prime = C \otimes I - (I + C)\otimes \ket{\vec 0}\bra{\vec 0}.
\end{equation}
Without loss of generality, we assume that the searched node is
labeled $\ket{\vec 0}$. The detailed analysis of this algorithm
shows that after $\frac{\pi}{2}\sqrt{2^{n-1}}=O(\sqrt{N})$
iterations, a measurement of the position of the walker yields the
marked state with success probability $\frac 12 - O(1/n)$
\cite{Shenvi}.

\section{AKR Algorithm}\label{sec:AKR}

The quantum search on a $\sqrt{N}\times\sqrt{N}$ grid has a
four-dimensional Hilbert space $\H_C$ for the coin and a
$N$-dimensional Hilbert space $\H_P$ for the $N$ lattice sites. A
basis for $\H_C$ is $\{\ket{d,j}\}$, for $0 \leq d \leq 1$ and $0
\leq j \leq 1$. Variable $d$ sets the direction of the walk, $d=0$
for horizontal shift and $d=1$ for vertical shift. Variable $j$
sets whether the walker moves forwards ($j=0$) or backwards
($j=1$).  A basis for $\H_P$ is $\{\ket{n_0,n_1}\}$, for $0 \leq
n_0,n_1
 \leq \sqrt{N}$. The boundary conditions are periodic.

The generic state of the quantum walker is
\begin{equation}
\ket{\Psi(t)} = \sum_{d,j=0}^{1}\sum_{n_0,n_1=0}^{\sqrt{N}}
\psi_{d,j;n_0,n_1}(t) \ket{d,j}\ket{n_0,n_1}.
\end{equation}
The action of the shift operator on the computational basis is
\begin{equation}
S\ket{d,j}\ket{n_0,n_1} = \ket{d,j\oplus 1
}\ket{n_0+(-1)^j\delta_{d,0},n_1+(-1)^j\delta_{d,1}},
\end{equation}
where $\oplus$ is the binary sum. Notice that there is an
inversion from backwards to forwards and vice-versa after the
action of the shift operator. This is a modification of the
standard shift operator and it is absolutely necessary to have a
quadratic speedup over the optimal classical algorithm.

The initial state is the uniform superposition of coin and spatial
eigenstates, {\it i.e.}, $\ket{\psi_0} = \ket{s^C}\otimes
\ket{s^P}$, where $\ket{s^P}$ is the uniform  superposition over
the $N$ sites of the grid and
\begin{equation}
\ket{s^C}=\frac{1}{2}\sum_{d,j=0}^{1}\ket{d,j}.
\end{equation}
The coin operator on unmarked vertices is given by
Eq.~(\ref{eq:grover-coin}) and the modified coin operator is
\begin{equation}
C^\prime = C \otimes I - (I + C)\otimes
\ket{n_0,n_1}\bra{n_0,n_1},
\end{equation}
where $(n_0,n_1)$ is the marked vertex. The evolution operator is
$U^\prime=S\ C^\prime$ and must be applied $O(\sqrt{N \log N})$
times. The overlap between the final state and the marked vertex
is $\Theta(1/\sqrt{\log N})$. In order to improve the probability
of finding the marked vertex, it is necessary $O(\sqrt{\log N})$
rounds of the algorithm yielding an overall complexity of
$O(\sqrt{N} \log N)$. A very recent paper \cite{Tulsi08} improved
the overall complexity to $O(\sqrt{N \log N})$.

\section{Results for SKW Algorithm}

\begin{figure}[ht]
\centering
\includegraphics[height=0.4\textwidth, angle=270]{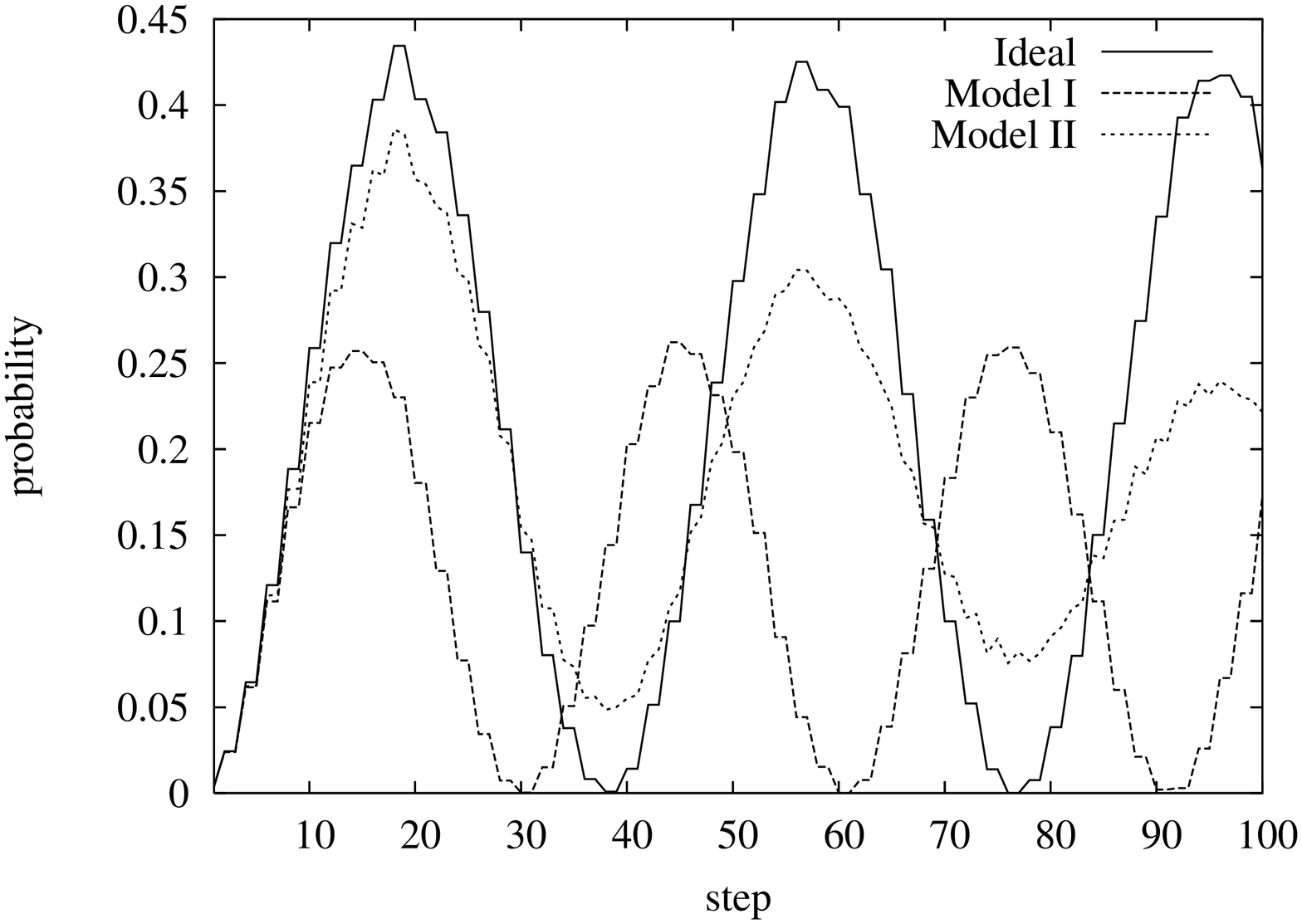}~
\includegraphics[height=0.4\textwidth, angle=270]{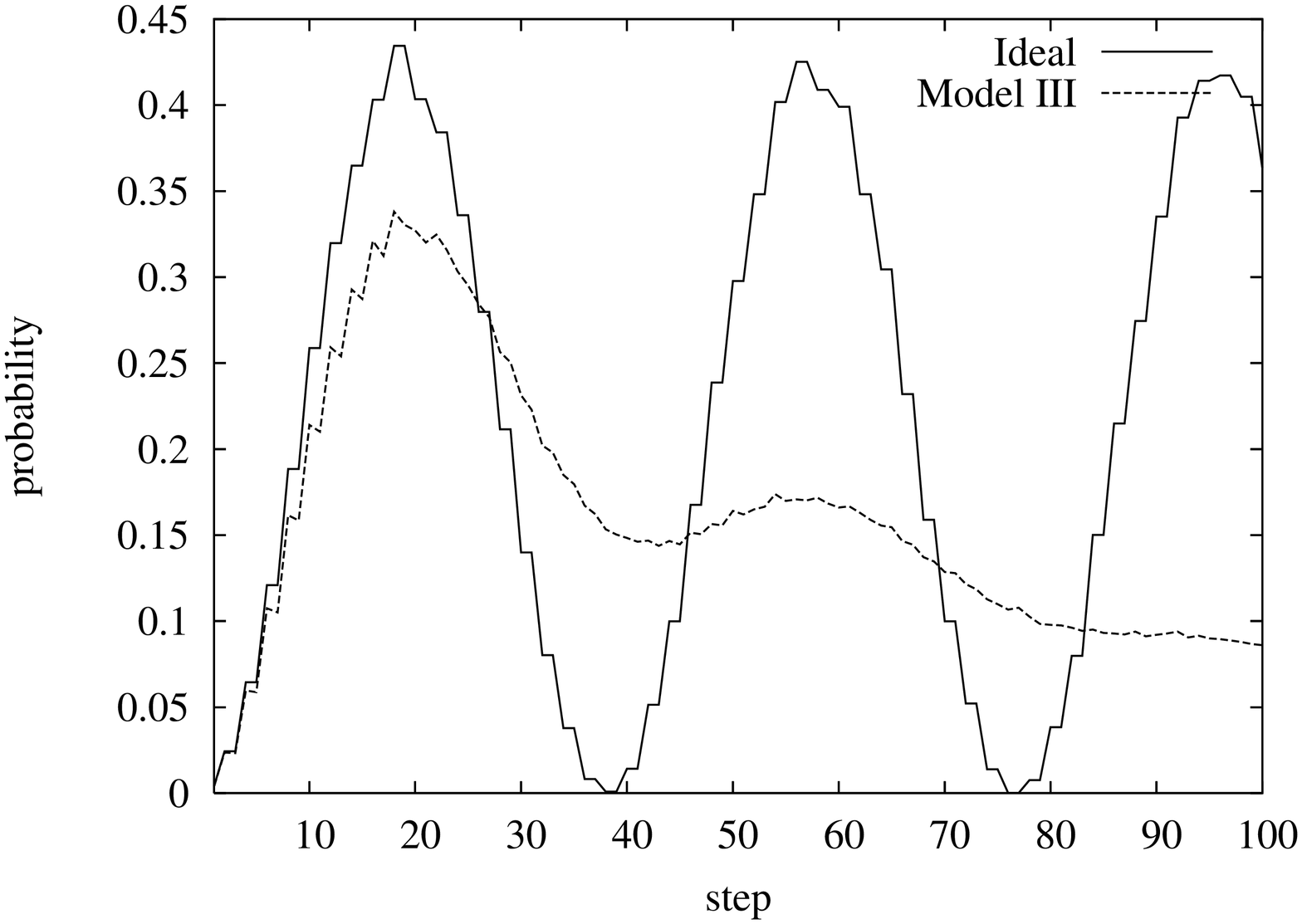}
\caption{Left panel: probability at the marked vertex as a function of the number of steps $s$ comparing the ideal case with both systematic ($\theta=0.3$) and random ($\sigma=0.3$) errors. Right panel: the same for broken-link errors with $p=0.02$.}
\label{fig:prob-vs-it}
\end{figure}

Fig.~\ref{fig:prob-vs-it} shows the probability of finding the
walker at the marked node as function of the number of steps, for
each noise model. In the left panel we compare the results for the
ideal case, without noise, with those for  noise models~I and~II.
In the right panel we compare the ideal case with the evolution
under noise model~III. All plots correspond to a hypercube of
dimension $n=8$. For model~I, we took a phase error $\theta=0.3$,
the standard deviation in model~II was $\sigma=0.3$ and the
probability of broken links per unit time (model~III) was
$p=0.02$. Note that the peak of probability for the systematic
error (model~I) occurs earlier than in the ideal case with zero
noise. The behavior of the algorithm under noise from model~I
affecting the coin operator at the marked node $(-I)$ is similar
to the one observed in \cite{Li2006}, in which the operator
$C$ was affected. Although random models~II and~III correspond to
different kinds of noise, they result in similar patterns. In both
cases the first maximum in the probability occurs approximately at
the same number of steps as in the case with no noise, and it
reaches a lower value. Subsequent peaks undergo a gradual
attenuation with the number of steps $s$.

\begin{figure}[ht]
\centering
\includegraphics[height=0.45\textwidth, angle=270]{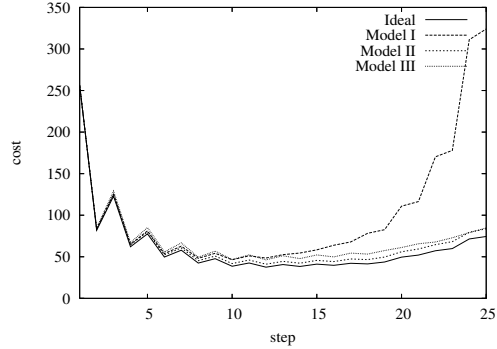}
\caption{Cost $c(s)$ from Eq.~(\ref{cost}), vs. number of steps,  for the noiseless search algorithm and for the algorithm with the three different noise models described in the text. The hypercube considered has dimension $n=8$.  }
\label{fig:cost}
\end{figure}

The stopping time in the case without noise corresponds to the
first maximum and this point is $\frac{\pi}{2}\sqrt{2^{n-1}}
\approx 18$, for $n=8$. In the presence of noise it is better to
stop the algorithm before this point and to rerun to find the
correct result. If the probability to obtain the correct result in
one run is $p$, then the expected number of times of trials is
$1/p$. If the computational complexity of one run is $O(\sqrt{N})$
then the overall complexity is $O(\sqrt{N}/p)$. If $p$ does no
depend on $N$, it is not going the change the complexity. Let us
define the algorithmic cost $c(s)$ as the overall cumulative
number of steps needed to find the desired state,
\begin{equation}\label{cost}
 c(s) = \frac{s}{p_s},
\end{equation}
where $s$ the number of steps before the final measurement is
taken in one run of the algorithm. In Fig.~\ref{fig:cost} we show
the cost function, $c(s)$, for the different noise models. In the
case of a systematic phase error, the cost function has a
well-defined minimum at $s\approx 10$. It is clearly convenient to
stop the algorithm before the peak probability is reached, in
either the ideal case or in the case with noise. For the other
noise models, and also in the case without noise, the cost
function has a very shallow minimum, and it increases very slowly
with step number after its minimum. However, even in these cases,
these  results suggest that it is advantageous to stop the
algorithm before the noiseless probability peak is reached, and to
repeat it more times, as needed.

\begin{figure}[ht]
\centering
\includegraphics[height=0.4\textwidth,angle=270]{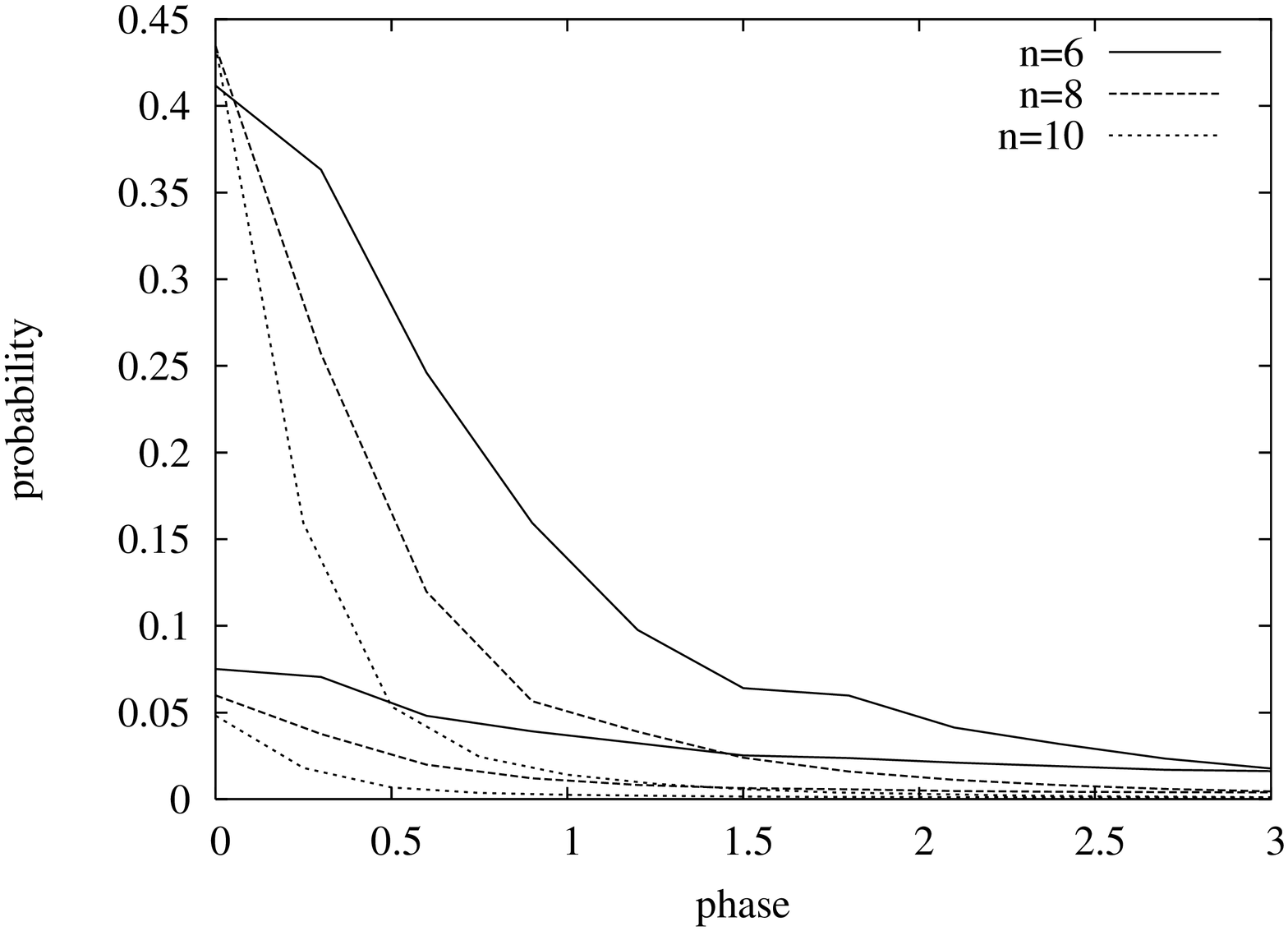}~
\includegraphics[height=0.4\textwidth,angle=270]{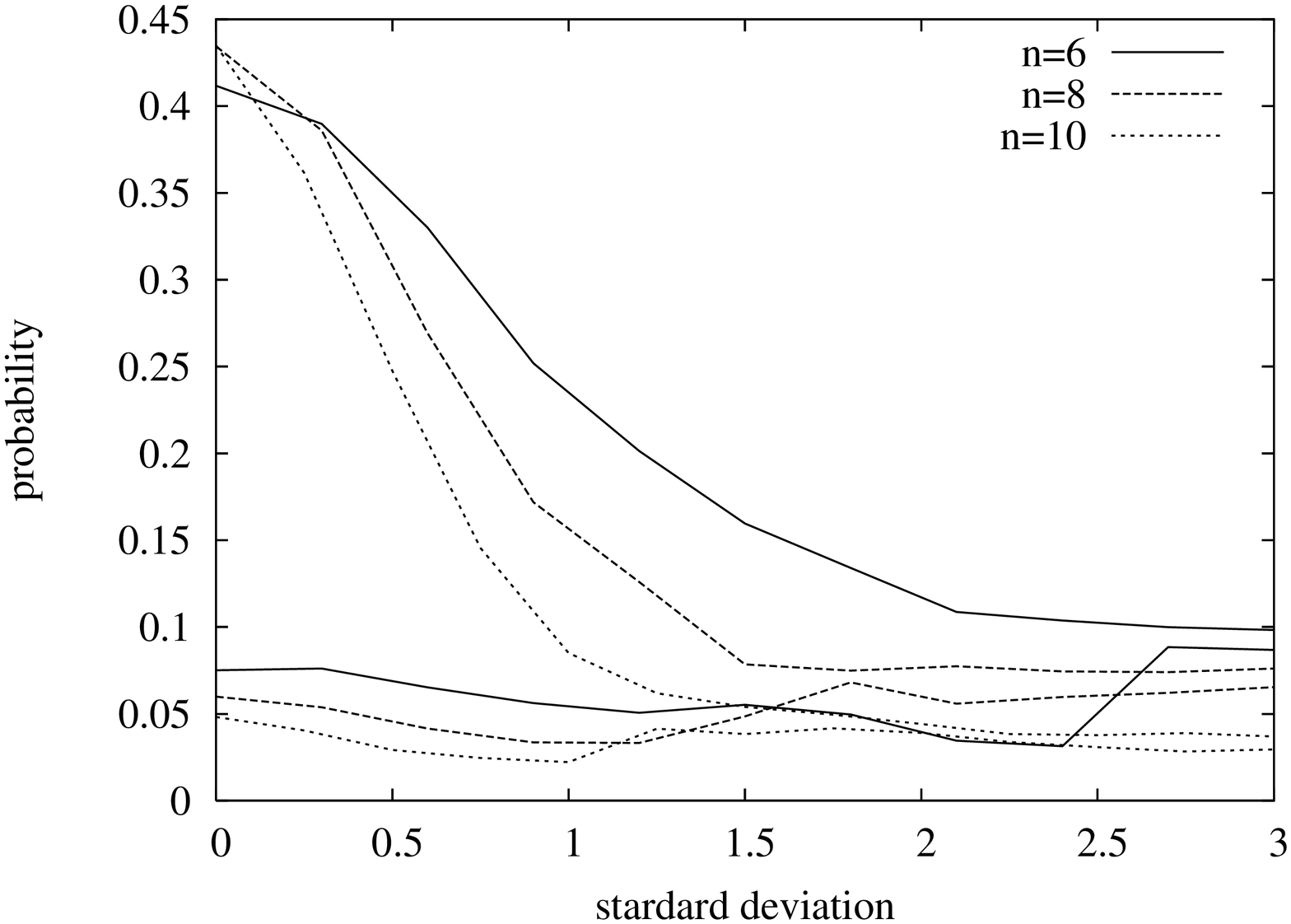}
\caption{Left panel: Results for model~I. Right panel: Results for model~II. Upper three curves: Highest probability at marked vertex as a function of the noise strength parameter for three values of the dimension of the hypercube $n$. Lower three curves: Highest probability at the unmarked vertices, using the same convention for the dependence of the dimension of the hypercube.}
\label{fig:probability-vs-noise}
\end{figure}

In Fig.~\ref{fig:probability-vs-noise} we observe the probability
of reaching the marked vertex as a function of the noise strength,
as represented by the three upper curves. In the left~(right)
panel we have the results for model~I~(model~II). Note that the
optimal phase is $\theta = 0$ or $\sigma=0$, i.e., when $-I$ is
used in the marked node as in the standard algorithm. The plot
also shows that the algorithm is very sensitive to noise from
operational errors, if no error correction code is used. The three
lower curves represent the highest probability among the unmarked
vertices. We observe that, as the phase error increases, the
difference between the maximum probabilities at marked and
unmarked nodes becomes smaller. In this case, we cannot
distinguish the right solution and the algorithm is no longer
useful.
The noise generated by systematic errors (left panel)
seems to play a more significant role in the algorithm than the noise generated by random errors (right panel).

\begin{figure}[ht]
\centering
\includegraphics[height=0.4\textwidth,angle=270]{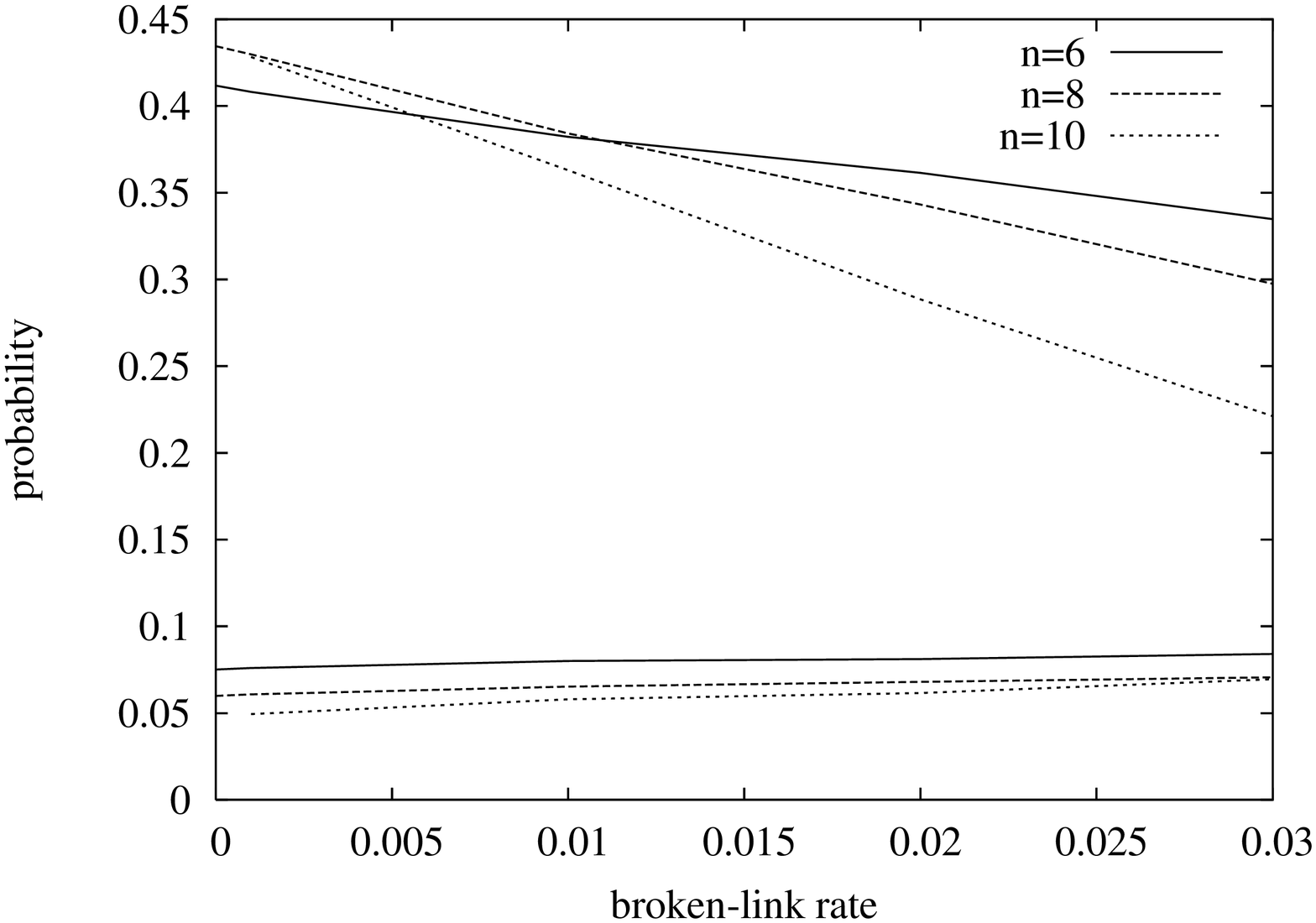}~
\includegraphics[height=0.4\textwidth,angle=270]{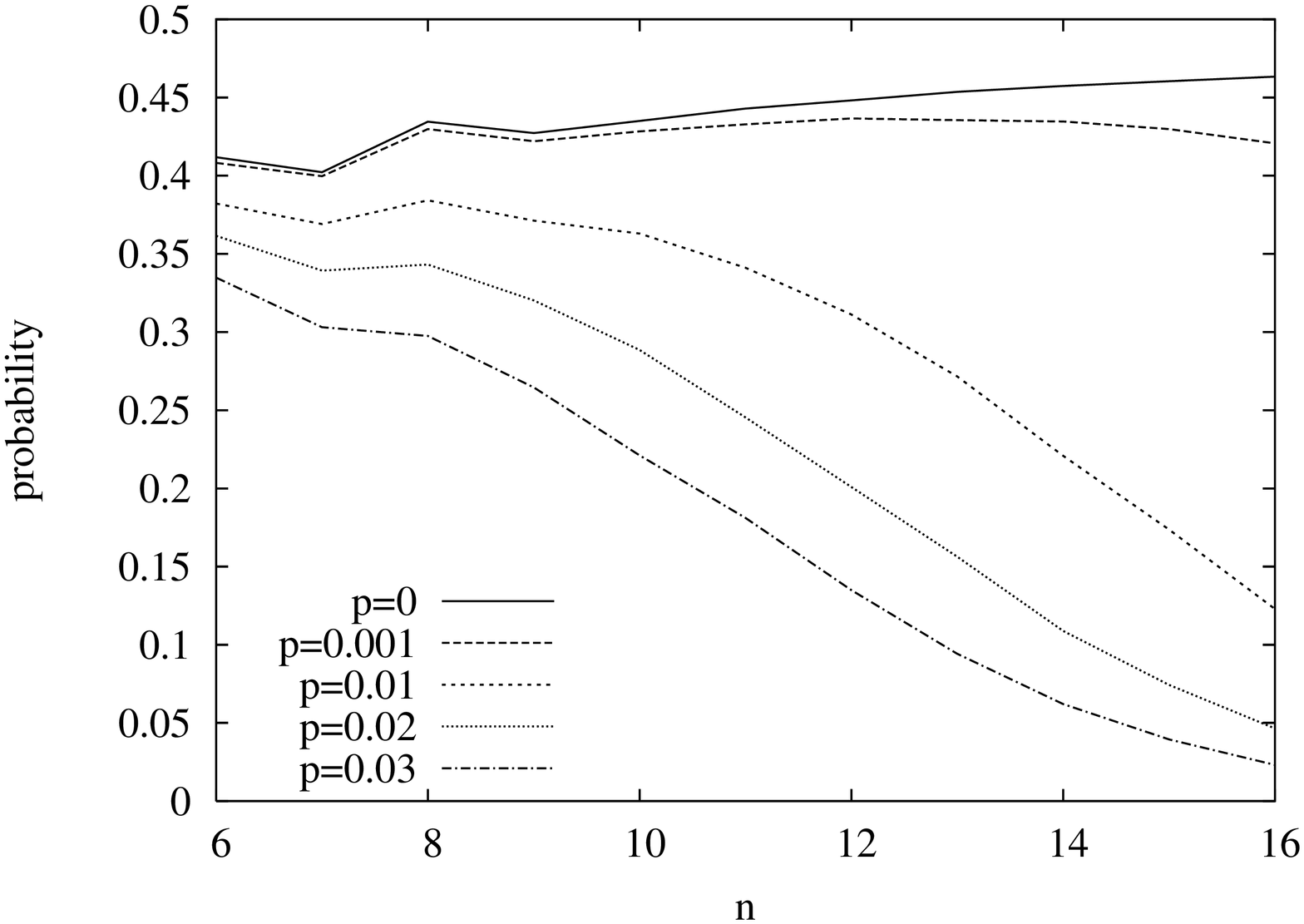}
\caption{Results for model III. Left panel: similar to
Fig.~\ref{fig:probability-vs-noise} as a function of the
broken-link rate. Right panel: highest probability at marked
vertex as a function of the dimension $n$ of the hypercube.}
\label{fig:probability-vs-blnoise}
\end{figure}

In Fig.~\ref{fig:probability-vs-blnoise} we have the results for
model~III. In the left panel, we observe the highest probability
at marked vertex as a function of the broken-link rate~$p$, as
represented by the three upper curves. The three lower curves
represent the highest probability among the unmarked vertices. In
the right panel, we have the highest probability at the marked
vertex as a function of the dimension $n$ of the hypercube. In
this case, the probabilities decay as the dimension of the
hypercube is increased, which is similar to the result obtained in
\cite{Li2006} for noise affecting the coin operator at
unmarked sites.

\begin{figure}[ht]
\centering
\includegraphics[width=0.5\textwidth]{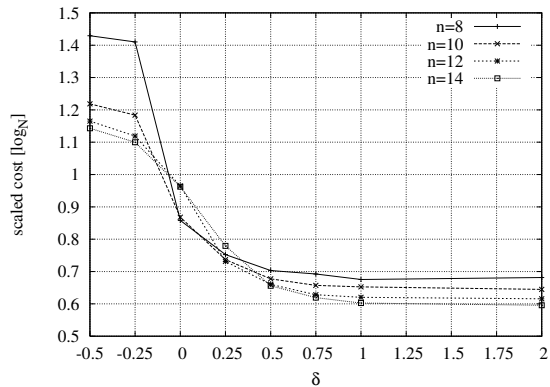}
\caption{Logarithm (base $N$) of the algorithmic cost as a function of
scaling parameter~$\delta$ for model~II comparing different dimensions.}
\label{fig:cost-vs-dim-err2}
\end{figure}

In order to estimate how errors of model~II change the complexity
of the algorithm we use a formula that scales with $N$ in the form
$\theta=1/N^\delta$ in Eq.~(\ref{Cv0}). In
Fig.~\ref{fig:cost-vs-dim-err2} we plot the scaled cost, which is
the logarithm to base $N$ of the algorithmic cost given by
Eq.~(\ref{cost}), against error parameter $\delta$ for several
values of $n$. Recall that the complexity of SKW algorithm is
$O(N^{0.5})$ and its success probability is $1/2-O(1/n)$. Hence,
for large values of~$\delta$ and~$N$, we should obtain a scaled
cost close to $0.5$, corresponding to the complexity of the
noiseless SKW algorithm. Our plot shows a scaled cost close to
$0.6$, which is consistent with the values of $N$ considered. This
means that for $\delta\geq 1$, the SKW algorithm with error has
the same complexity of the noiseless SKW algorithm. For
$\delta<1$, the noise rate increases and the algorithm gradually
looses efficiency in relation to the noiseless search. For
$\delta\approx -0.1$, the scaled cost is close to~$1$, which means
that the quantum algorithm has the same complexity of the
classical search, $O(N)$. For $\delta<-0.1$, the scaled cost is
higher than $1$, which means that the performance of the quantum
search is worse than the classical search.

\section{Results for AKR Algorithm}

The behaviour of the maximum probability at the marked node in the
AKR algorithm follows a similar pattern as in the
SKW algorithm. The main difference being that for AKR the maximum
probability decreases as $N$ increases, while for SKW the maximum
probability remains close to $1/2$. The numerical results for the
cost in the AKR algorithm also show that in the presence of imperfections it is better to stop the
algorithm before the theoretical stopping time.

\begin{figure}[ht]
\centering
\includegraphics[height=0.31\textwidth]{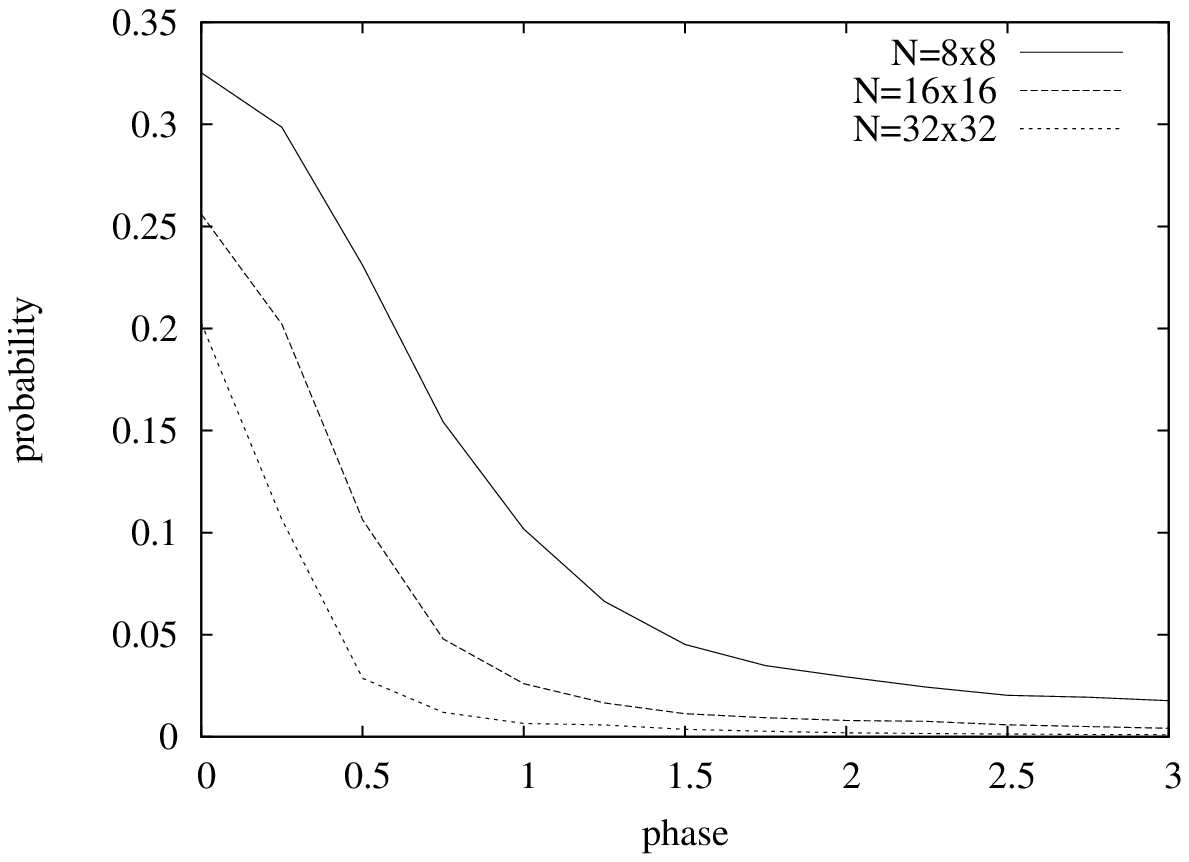}~
\includegraphics[height=0.31\textwidth]{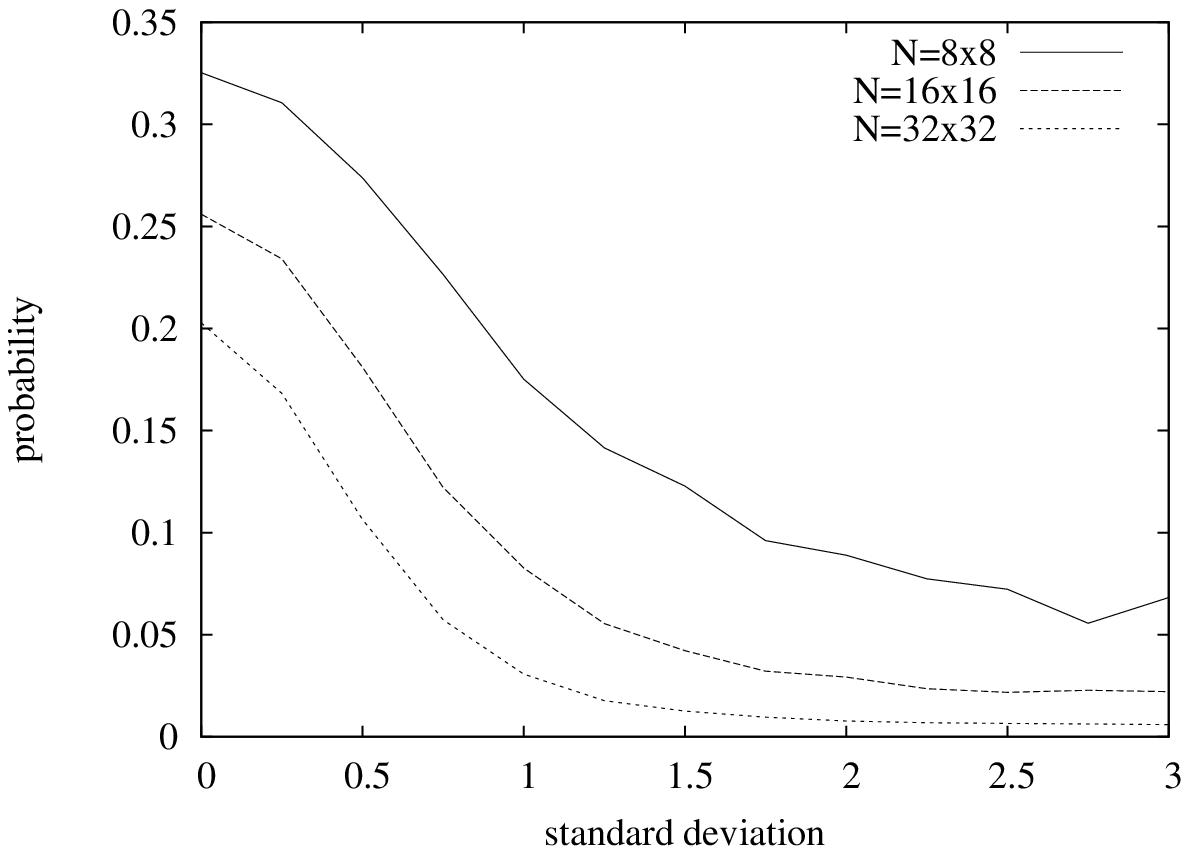}
\caption{Left panel: Results for model~I. Right panel: Results for
model~II. Highest probability at marked vertex as a function of
the noise strength parameter for three values of the dimension of
the grid.} \label{fig:probability-vs-noise-AKR}
\end{figure}

Fig.~\ref{fig:probability-vs-noise-AKR} shows the maximum
probability at the marked node as function of the noise strength.
In the left~(right) panel we have the results for
model~I~(model~II). This figure should be compared to
Fig.~\ref{fig:probability-vs-noise}.  Note that the number of
nodes of the grids corresponds to the number of nodes of the
hypercubes. The curves for the AKR algorithm are very similar to those for the SKW
algorithm and we draw similar conclusions for both cases. The main difference is
the distance between the curves, a consequence of the fact
that in the AKR algorithm the maximum probability at the marked node
drops when we increase $N$.

\begin{figure}[ht]
\centering
\includegraphics[height=0.31\textwidth]{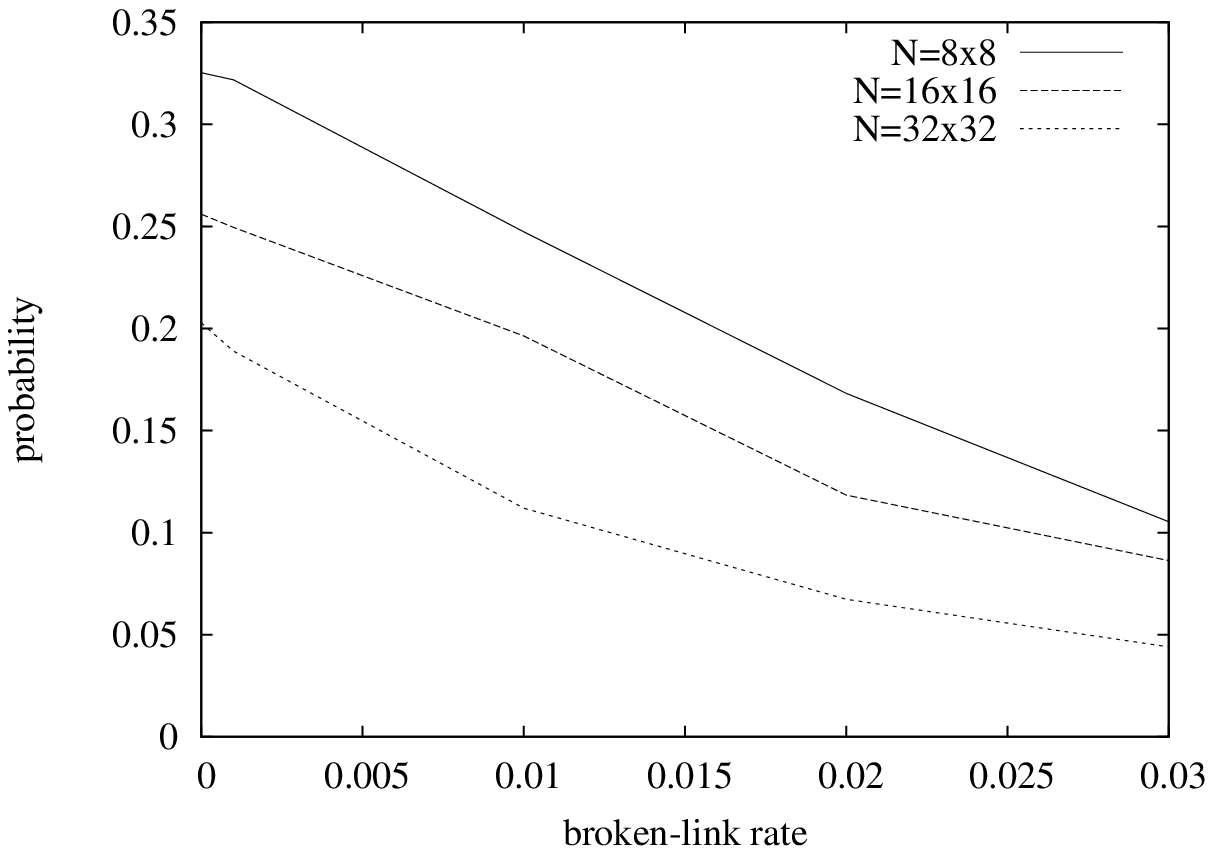}~
\includegraphics[height=0.31\textwidth]{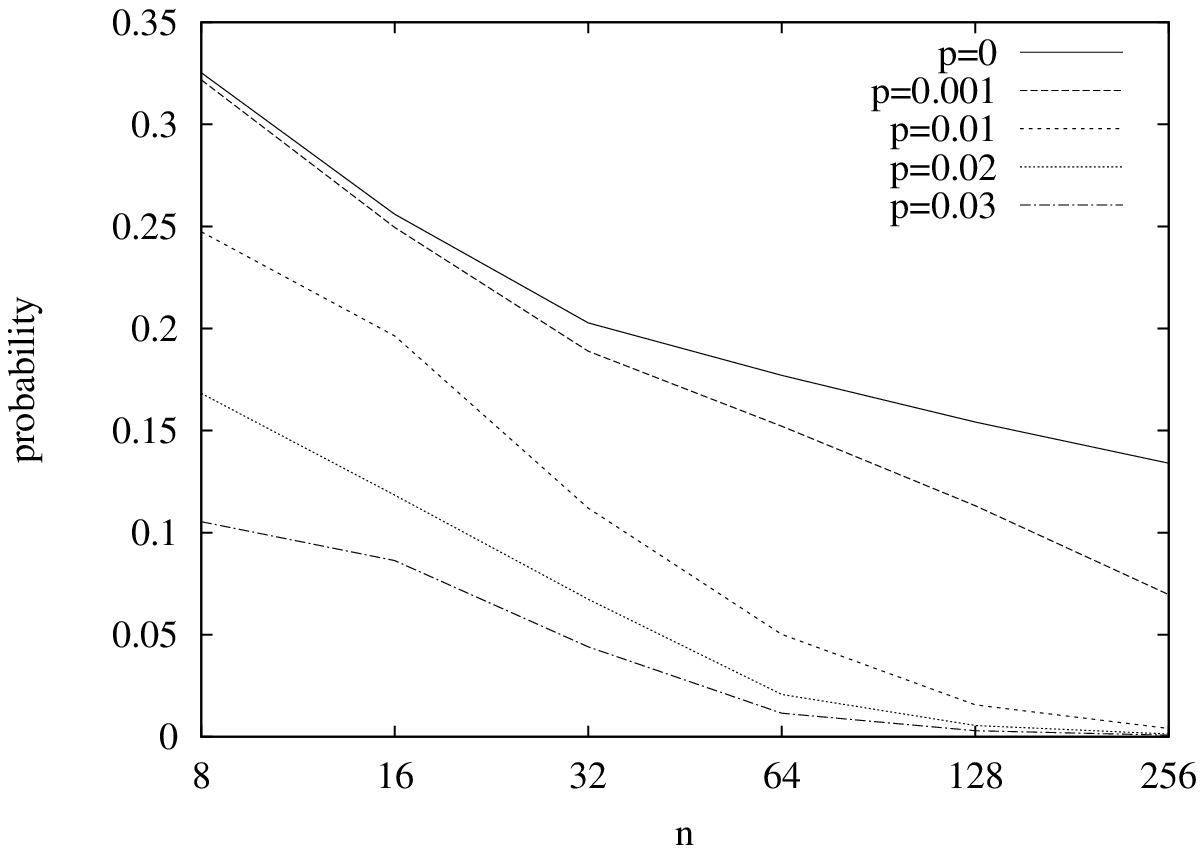}
\caption{Results for model III. Left panel: highest probability as
a function of the broken-link rate. Right panel: highest
probability at marked vertex as a function of the dimension $\log
N$ of the $\sqrt{N}\times\sqrt{N}$ grid.}
\label{prob-vs-rate-err3}
\end{figure}

In Fig.~\ref{prob-vs-rate-err3} we show the results for model~III.
In the left panel, we observe the highest probability at the marked
vertex as a function of the broken-link rate~$p$.
In the right panel, the horizontal axis is in log scale.
These results should be compared with
Fig.~\ref{fig:probability-vs-blnoise}.  The
probability drops faster in the AKR algorithm than in the SKR. This was
predicted in \cite{AKR}, where it is shown that the probability at
the marked node scales as $O(1/\sqrt{\log N})$.

\begin{figure}[ht]
\centering
\includegraphics[width=0.5\textwidth]{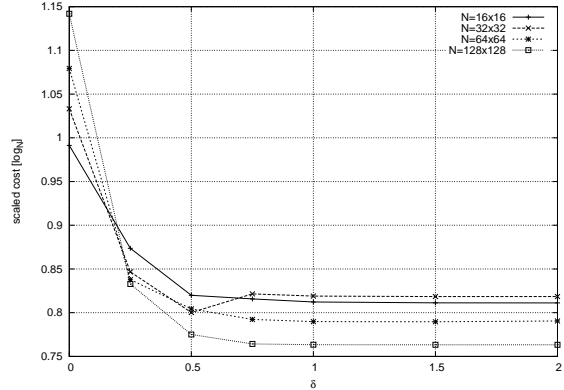}
\caption{Logarithm (base $N$) of the algorithmic cost as a
function of scaling parameter~$\delta$ for model~II comparing
different dimensions in AKR algorithm.}
\label{fig:cost-vs-dim-AKR}
\end{figure}

In Fig.~\ref{fig:cost-vs-dim-AKR} we plot the scaled cost $\log_N
c(s)$ against $\delta$, which is the analogue of
Fig.~\ref{fig:cost-vs-dim-err2} for the AKR algorithm. Recall that
the cost in the AKR algorithm is $O(N^{0.5}\log N)$. Hence, for large
values of~$\delta$ and~$N$, we should obtain a scaled cost a
little bit above $0.5$, corresponding to the complexity of the
noiseless AKR algorithm. The scaled cost is not exactly the power
of $N$ because the cost has the term $\log N$. Our plot shows a
scaled cost close to $0.8$, which is consistent with the values of
$N$ considered. From the figure we see that for $\delta\geq 1/2$,
the AKR algorithm with error has the same complexity as the
noiseless AKR algorithm. For $\delta<1/2$, the noise rate
increases fast enough such that the algorithm looses efficiency in
relation to the noiseless search. When we decrease $\delta$, the
scaled cost approaches 1, which means that the quantum algorithm
has the same complexity of the classical search, $O(N)$. For
$\delta<0$, the scaled cost is higher than $1$, which means that
the performance of the quantum search is worse than the classical
search. Note that $\delta=1/2$ is the transition point in the AKR
algorithm, while $\delta=1$ is the transition point in the SKW
algorithm. For comparison, note that \cite{Shenvi-noise} obtained
$\delta=1/4$ as the transition point in the original Grover's
algorithm.

\section{Conclusions}

In this work, the effect of imperfect quantum operators on search
algorithms based on quantum walks has been investigated. We have
considered both systematic and random phase errors in the coin
operator. The effect of randomly broken links affecting the shift
operator has also been considered. This kind of error directly
affects the spatial propagation of the walker. We have considered
the search of a marked vertex on hypercubes (SKW algorithm) and
two-dimensional grids (AKR algorithm).

For the SKW algorithm, we found that the overall effect of
noise in the coin operator for the marked node is similar to
that for the imperfect coin operator acting on unmarked nodes
considered in \cite{Li2006}. There are also many similarities
with the AKR algorithm. The overall qualitative effect of noise
seems to be similar in all search algorithms considered. On the other hand, we
have obtained quantitative results for the tolerance of the
algorithms to errors.

In the context of  Grover's search algorithm, it was shown
analytically that phase errors $\theta$ scaling as $1/N^\delta$,
for $\delta \le 1/4$, modify the complexity of the algorithm to
$O(N^{1-2\delta})$ \cite{Shenvi-noise}. For $\delta\geq 1/4$, the
complexity of Grover's algorithm with errors is equal to the
complexity of the noiseless case, $O(\sqrt{N})$. If $\delta
<1/4$, the advantage over a classical search, $O(N)$, is
progressively reduced. If $\delta=0$, \textit{i.e.}, constant
error, the complexity of Grover's algorithm is equal to the
complexity of the classical search. We have also considered this
setup in the context of the SKW and AKR search algorithms. Our
numerical simulations show that the transition point for AKR is
around $\delta=1/2$ and for SKW it is around $\delta=1$. For $\delta$
below these threshold values, the algorithm gradually looses
efficiency until becoming worse than the classical case at around
$\delta=0$.

Our numerical results show that it is possible to improve the
efficiency in all cases (with or without noise) if we stop the
search algorithm before the number of steps predicted
theoretically. In this case, more than one round of the algorithm
is needed, keeping the overall cost smaller than when using the
theoretical stopping point.

\paragraph{Acknowledgments.} F.L.M., A.C.O. and R.P. acknowledge financial support of
CNPq. G.A. and R.D. acknowledge support from PEDECIBA-Uruguay. We
acknowledge support from \textit{Edital CT-INFO} n.07/2007 ---
\textit{Grandes Desafios da Ci\^{e}ncia da Computa\c{c}\~{a}o no Brasil
2006-2016}.


\begin{thebibliography}{10}

\bibitem[Alagic and Russell (2005)]{Alagic}
G.~Alagic and A.~Russell, \emph{Decoherence in quantum walks on the hypercube},
  Phys. Rev. A \textbf{72} (2005), 062304.

\bibitem[Ambainis (2004)]{Amb03}
A.~Ambainis, \emph{Quantum walk algorithm for element distinctness},
  Proceedings 45th Annual IEEE Symp. on Foundations of Computer Science (FOCS),
  2004.

\bibitem[AKR (2005)]{AKR} 
Andris Ambainis, Julia Kempe, and Alexander Rivosh, \emph{Coins make quantum
  walks faster}, SODA '05: Proceedings of the sixteenth annual ACM-SIAM
  symposium on Discrete algorithms (Philadelphia, PA, USA), Society for
  Industrial and Applied Mathematics, 2005, pp.~1099--1108.

\bibitem[Grover (1996)]{Gro96a}
L.~Grover, \emph{A fast quantum mechanical algorithm for database search},
  Proc. 28th Annual ACM Symposium on the Theory of Computation (New York, NY),
  ACM Press, New York, 1996, pp.~212--219.

\bibitem[Kempe (2003)]{Kempe03}
J.~Kempe, \emph{Quantum random walks -- an introductory overview}, Contemp.
  Phys. \textbf{44} (2003), no.~4, 307--327.

\bibitem[Kendon and Tregenna (2003)]{KT02}
V.~Kendon and B.~Tregenna, \emph{Decoherence can be useful in quantum walks},
  Phys. Rev. A \textbf{67} (2003), 042315.

\bibitem[Li, Ma, and Zhou (2006)]{Li2006}
Yun Li, Lei Ma, and Jie Zhou, \emph{Gate imperfection in the quantum
  random-walk search algorithm}, Journal of Physics A \textbf{39} (2006),
  9309--9319.

\bibitem[Long, \textit{et. al.} (2000)]{Long00}
Gui~Lu Long, Yan~Song Li, Wei~Lin Zhang, and Chang~Cun Tu, \emph{Dominant gate
  imperfection in grover\char39{}s quantum search algorithm}, Phys. Rev. A
  \textbf{61} (2000), no.~4, 042305.

\bibitem[Magniez, Santha and Szegedy (2007)]{mss07}
F.~Magniez, M.~Santha, and M.~Szegedy, \emph{Quantum algorithms for the
  triangle problem}, SIAM Journal on Computing \textbf{37} (2007), no.~2,
  413--424.

\bibitem[Marquezino, Portugal, Abal and Donangelo (2008)]{Mixing}
F.~L. Marquezino, R.~Portugal, G.~Abal, and R.~Donangelo, \emph{Mixing times in
  quantum walks on the hypercube}, Physical Review A \textbf{77} (2008),
  042312.

\bibitem[Moore and Russell (2002)]{Moore}
C.~Moore and A.~Russell, \emph{Quantum walks on the hypercube}, Proceedings of
  6th International Workshop on Randomization and Approximation Techniques
  (RANDOM 2002), Vol. 2483 of Lecture Notes in Computer Science (LNCS)
  (Cambridge, MA) (J.~D.~P. Rolim and S.~Vadhan, eds.), Springer-Verlag,
  Berlin, 2002, 2002, pp.~164--178.

\bibitem[Nielsen and Chuang (2000)]{NC00}
Michael~A. Nielsen and Isaac~L. Chuang, \emph{Quantum computation and quantum
  information}, {Cambridge University Press}, October 2000.

\bibitem[Oliveira, Portugal and Donangelo (2006)]{Amanda}
A.C. Oliveira, R.~Portugal, and R.~Donangelo, \emph{Decoherence in
  two-dimensional quantum walks}, Phys. Rev. A \textbf{74} (2006), 012312.

\bibitem[Romanelli, \textit{et. al.} (2004)]{deco}
A.~Romanelli, R.~Siri, G.~Abal, A.~Auyuanet, and R.~Donangelo,
  \emph{Decoherence in the quantum walk on the line}, Physica A \textbf{347}
  (2004), 137--152.

\bibitem[Shenvi, Brown and Whaley (2003)]{Shenvi-noise}
N.~Shenvi, K.~R. Brown, and K.~B. Whaley, \emph{Effects of a random noisy
  oracle on search algorithm complexity}, Physical Review A \textbf{68} (2003),
  052313.

\bibitem[SKW (2003)]{Shenvi} 
N.~Shenvi, J.~Kempe, and K.~B. Whaley, \emph{A quantum random walk search
  algorithm}, Physical Review A \textbf{67} (2003), 052307.

\bibitem[Szegedy (2004)]{Szegedy04}
Mario Szegedy, \emph{Quantum speed-up of markov chain based algorithms},
  Foundations of Computer Science, Annual IEEE Symposium on \textbf{0} (2004),
  32--41.

\bibitem[Tulsi (2008)]{Tulsi08}
A.~Tulsi, \emph{Faster quantum walk algorithm for the two dimensional spatial
  search}, Phys. Rev. A \textbf{78} (2008), no.~1, 012310.

\end{thebibliography}
\end{document}